\documentclass[letter]{aa}
\usepackage{amsmath}
\usepackage{subfigure}
\usepackage[T1]{fontenc}
\usepackage{booktabs}
\usepackage{txfonts}
\usepackage{sistyle}
\usepackage{tablefootnote}
\usepackage{color}
\usepackage{graphicx}
\usepackage{mhchem}
\usepackage[normalem]{ulem} % Hugo, to strike out

\begin{document}

\title{Formation of the Trappist-1 system in a dry protoplanetary disk}

\author{Antoine Schneeberger\inst{1}, Olivier Mousis\inst{1,2}, Magali Deleuil\inst{1,2}, Jonathan I. Lunine\inst{3} }

\institute{Aix- Marseille Universit\'e, CNRS, CNES, Institut Origines, LAM, Marseille, France \\\email{antoine.schneeberger@lam.fr}
\and Institut Universitaire de France (IUF), France
\and Department of Astronomy, Cornell University, Ithaca, NY, USA}

\authorrunning{A. Schneeberger et al.}

\date{Received 18/10/2023  /
Accepted 18/01/2024 }

\abstract{A key feature of the Trappist-1 system is its monotonic decrease in bulk density with growing distance from the central star, which indicates an ice mass fraction that is zero in the innermost planets, b and c, and about 10\% in planets d through h. Previous studies suggest that the density gradient of this system could be due to the growth of planets from icy planetesimals that progressively lost their volatile content during their inward drift through the protoplanetary disk. Here we investigate the alternative possibility that the planets formed in a dry protoplanetary disk populated with pebbles made of phyllosilicates, a class of hydrated minerals with a water fraction possibly exceeding 10 wt\%. We show that the dehydration of these minerals in the inner regions of the disk and the outward diffusion of the released vapor up to the ice-line location allow the condensation of ice onto grains. Pebbles with water mass fractions consistent with those of planets d--h would have formed at the snow-line location. In contrast, planets b and c would have been accreted from drier material in regions closer to the star than the phyllosilicate dehydration line.}


\maketitle

\section{Introduction} \label{sec:intro}

Because of its compactness and density gradient, the Trappist-1 exoplanetary system is the closest analog to the well-characterized Galilean moon system. It is also an outstanding laboratory for exploring planet formation around ultracool M dwarfs. The Trappist-1 system contains seven exoplanets in a resonance chain, three of which are  within the habitable zone \citep{gil16,gil17}, with semimajor axes ranging between 0.015 and 0.062 AU \citep{lug17,ago21}. This resonance chain could be a sign of a past inward drift of the whole system before orbital capture \citep{orm17,bit19,col19,izi21,mad21}. 

The measured densities of the Trappist-1 planets are among the most accurate ones available for an exoplanetary system \citep{lug17}. These densities have been used by \cite{acu21} to assess their interiors and compositions. The authors show evidence for an increasing water mass fraction (WMF) of the planets with increasing distance from the host star. The planets can be separated into two groups, with planets b--c containing little to zero water, and planets d--h presenting significant amounts of water (see Table \ref{tab:acuna}). The significant amount of water estimated in the five outermost planets has been interpreted as a sign that they formed beyond the position of the snow line (SL), within the water-rich protoplanetary disk (PPD), prior to their inward migration \citep{orm17,joh22}. Such a mechanism requires a rapid growth of the planets, which could be eased by the formation of planetesimals via streaming instability triggered by a solid-to-gas ratio above unity at the SL due to ice enrichment \citep{orm17}. 

However, red dwarfs produce far-ultraviolet (FUV) and extreme-ultraviolet (XUV) irradiation during their early evolution \citep{shk14}. This can cause the photolytic destruction of the water present in the PPD or in the material originating from the interstellar medium and accreted by the PPD \citep{shk14,loy18,you21}. Furthermore, repeated episodes of accretional outbursts by Trappist-1 could have produced rapid temperature rises in the PPD that would  dehydrate pebbles \citep{hou23}. {Finally}, sequestration of oxygen in rock- and metal-forming species must be considered when the carbon-to-oxygen (C/O) ratio exceeds the solar value (0.55; \citealt{lod21}) in the PPD.  When the C/O ratio reaches or exceeds 0.8, the depletion of oxygen is severe; it is bound up in rock, metal oxides, CO, and CO$_2$, hindering the production of water \citep{pek19}.

In systems where one or both of the water loss processes occur, an interesting alternative is to consider the case where water would be embedded in phyllosilicate particles. Phyllosilicates form a large group of hydrous minerals, which are structured in layers bounded by cation--anion couples that contain OH groups in their crystalline structure. They can contain H$_2$O molecules between layers, and dehydration experiments show that serpentinite minerals (one of the most common forms) can hold up to $\sim$13 wt$\%$ of water \citep{bez10,cha21,pet22}

Detection of phyllosilicate grains was claimed in the interstellar medium and PPDs \citep{zai75,whi97,rea09}. However, this evidence has since been reinterpreted as a mixture of amorphous silicates and water ice \citep{gib04, pot21}. On the other hand, phyllosilicates are found in CI and CM chondrites, which have WMFs as high as 10\% \citep{bec10,bec14,ale19}. The current consensus is that phyllosilicates found in CM chondrites  are formed by aqueous alteration of silicates in water-rich planetesimals \citep{bis88,sut21}.  Such ice-rich solids can form in the vicinity of the water SL and become precursors of phyllosilicate-rich pebbles \citep{bis88} via collisional evolution.

However, hydrous mineral inclusions within chondritic meteorites exhibit a distinct D/H ratio and hydration heterogeneity, favoring a primitive origin \citep{met92,cie03,pia21}. This supports models that show that micrometric phyllosilicate grains can result from the alteration of silicate grains by water vapor {over a time period of $\sim$100 kyr in the PPD} \citep{ang19,thi20}. {This is also consistent with the idea that phyllosilicate nanocrystals could form in the wake of shock waves crossing the PPD \citep{cie03}.
These shock waves would produce a temperature rise of up to 2000 K \citep{cie03,bur19}, which would evaporate water ice and melt silicates.  As the molten silicate recrystallized, it would react with the surrounding water vapor to form phyllosilicate nanocrystals within a few days of the shock wave. }{Both processes require the presence of water vapor, which implies that they occurred early in the history of the PPD.} 

\begin{figure*}[h]
\centering
\includegraphics[width=\linewidth, trim={0 2cm 0 1.5cm}, clip]{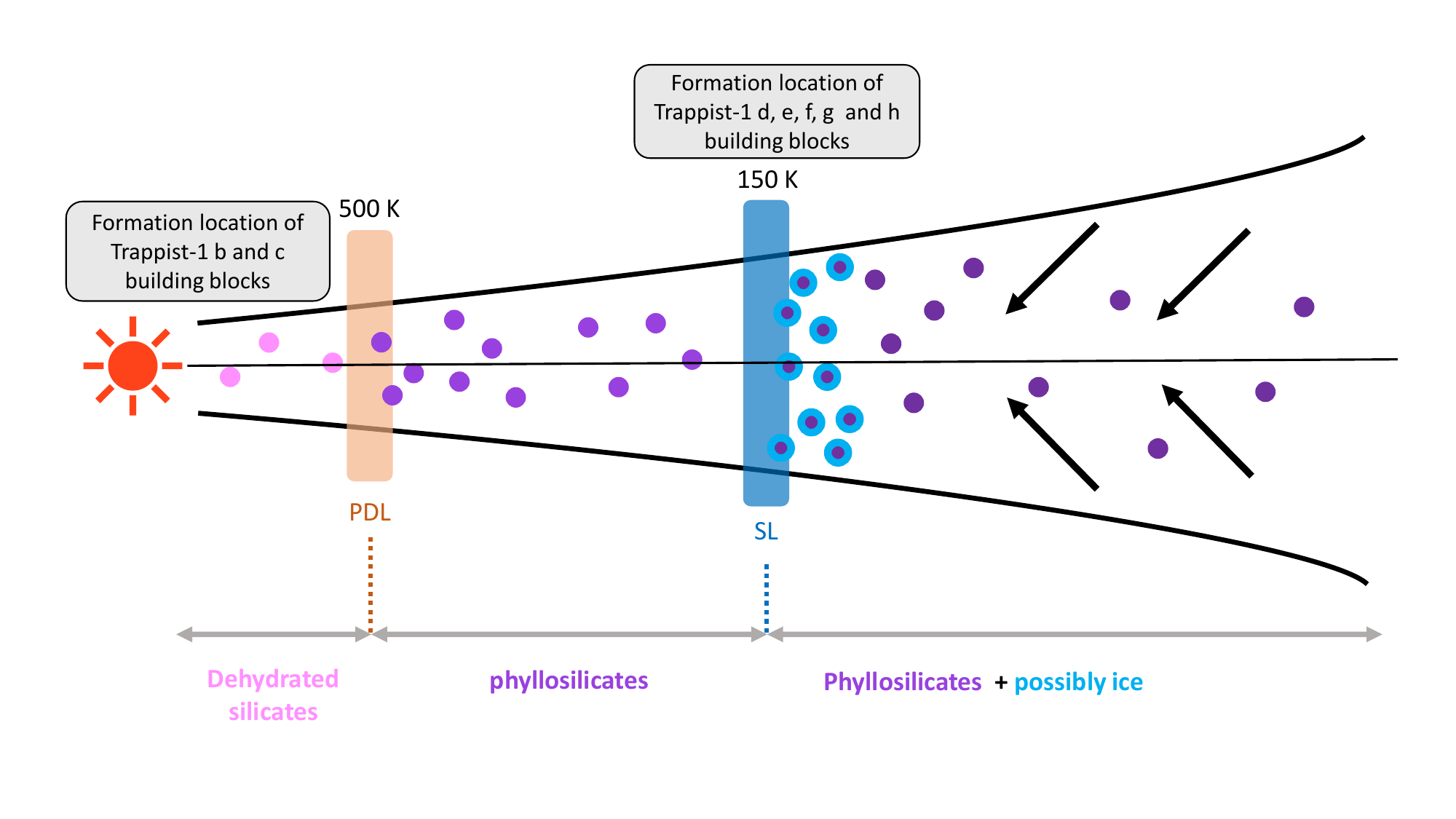} 
\caption{Structure of Trappist-1's PPD. Phyllosilicate grains (purple dots) drift inward in the disk. When these grains cross the PDL (orange bar), they dehydrate and release water as vapor. Only silicate grains remain inside the PDL (pink dots). The vapor viscously diffuses inward and outward and eventually crosses the SL (blue bar), where it condenses and coats phyllosilicate grains with ice (blue shell).}
\label{fig:sketch}
\end{figure*}

The possible dearth of water as ice in red dwarf PPDs suggests an alternative scenario where (i) the Trappist system formed in a dry PPD devoid of volatiles, and (ii) water was only delivered to the inner disk in the form of phyllosilicate minerals. Figure \ref{fig:sketch} qualitatively describes this scenario in which three distinct regions coexist in the disk: an inner region occupied by dehydrated phyllosilicates that is located inside the phyllosilicate dehydration line (PDL), a second region containing hydrated phyllosilicates and bracketed by the PDL and the SL, and an outer region populated by phyllosilicates coated with ice and extending from the SL. In this scenario, phyllosilicates dehydrate in the inner part of the disk due to increasing temperature and release water vapor at the PDL, which viscously diffuses inward and outward, eventually crossing the SL, where it condenses onto grains. This formation mechanism has been successfully applied to the Galilean moons to account for their observed density gradient \citep{mou23}. 

 We have investigated the evolution of the WMF of phyllosilicate-rich pebbles during their transport and growth in a dry PPD. We used the one-dimensional volatile evolution model in a PPD from \cite{agu20,agu22} and \cite{sch23}, to which we added the dehydration of phyllosilicates at the PDL location. We tracked the WMF evolution of pebbles at the SL location, assuming a starting WMF in the phyllosilicate grains in the 1--10\% range. We then studied the impact of the proposed scenario on the formation history of the Trappist-1 system.

\begin{table}[h]
\centering
\caption{Estimated masses and WMFs for Trappist-1 planets \citep{acu21}.} 
\begin{tabular}{@{}lcc@{}}
\hline
\hline           
\smallskip
\smallskip
Planet                      &  Mass (M$_{\oplus}$)            & WMF \\ 
\hline           
\smallskip
b   &   $1.375 \pm 0.041$   &   $(3.1^{+0.5}_{-3.1}) \times 10^{-5}$ \\
c   &   $1.300 \pm 0.036$   &   $(0.0^{+4.4}_{-0.0}) \times 10^{-6}$ \\
d   &   $0.388 \pm 0.007$   &   $0.084 \pm 0.071$                    \\ 
e   &   $0.699 \pm 0.013$   &   $0.094 \pm 0.067$                    \\ 
f   &   $1.043 \pm 0.019$   &   $0.105 \pm 0.073$                    \\
g   &   $1.327 \pm 0.024$   &   $0.119 \pm 0.080$                    \\
h   &   $0.327 \pm 0.012$   &   $0.081^{+0.059}_{-0.081}$            \\
\hline           
\end{tabular}
\label{tab:acuna}
    \end{table}

\section{Disk model}

Our PPD model is based on the PPD model presented in \cite{agu20}, \cite{mou20}, \cite{agu22}, and \cite{sch23}. To size the model to the Trappist-1 system, the central star mass was scaled down to 0.0802 $M_\odot$ \citep{gil16}. We also opted to linearly scale the disk mass and accretion rate from 0.1 $M_\odot$ and  $10^{-7.6}$ $M_\odot$yr$^{-1}$ for the PPD \citep{har98} to 0.1 $M_\star$ and  $10^{-7.6}$ $M_\star$yr$^{-1}$ for the Trappist-1 system, with $M_\star$ being the mass of Trappist-1.

Our PPD model follows the differential equation \citep{lyn74}

\begin{equation}
\frac{\partial \Sigma_{\mathrm{g}}}{\partial t} = \frac{3}{r} \frac{\partial}{\partial r} \left[ r^{1/2} \frac{\partial}{\partial r} \left( r^{1/2} \Sigma_{\mathrm{g}} \nu \right) \right],
\label{eq:psn}
\end{equation}

\noindent where $\Sigma_{\mathrm{g}}$ is the gas surface density and $\nu$ is the gas viscosity. Our model assumes invariance in the orbital direction and hydrostatic equilibrium in the azimuthal direction. The viscosity is given by \citep{sha73}
\begin{equation}
\nu = \alpha \frac{C_s^2  }{\Omega_k}
,\end{equation}

\noindent where $\alpha$ is the viscosity parameter set to the canonical value of $10^{-3}$ \citep{sha73,lyn74}, $C_s$ the sound of speed, and $\Omega_k$ the Keplerian pulsation. 

The initial disk profile was computed from the self-similar solution derived by \cite{lyn74}:

\begin{equation}
\Sigma_{\mathrm{g},0} = \frac{\dot{M}_{\mathrm{acc},0}}{3 \pi \nu } \exp \left[ - \left( \frac{r}{r_\mathrm{c}} \right)^{0.5} \right] 
\label{eq:init}
,\end{equation}

\noindent with $\dot{M}_{\mathrm{acc},0}$ the initial accretion rate onto the PPD and $r_\mathrm{c}$ the centrifugal radius of the disk. The centrifugal radius was self-consistently computed at its initialization as described in \cite{agu20}. 

\begin{figure}[h]
\centering
\includegraphics[width=0.9\columnwidth]{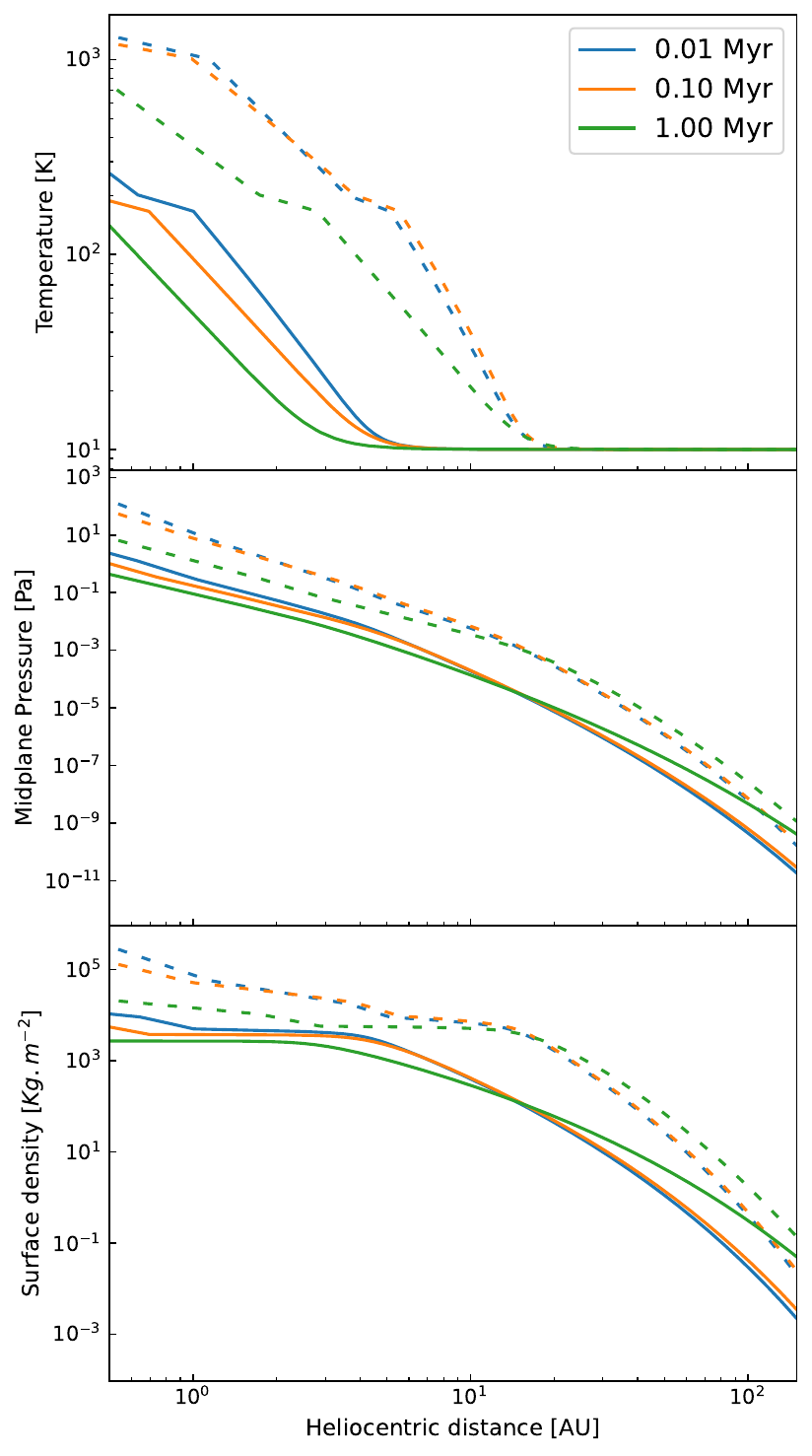}
\caption{From top to bottom: Temperature, pressure, and surface density profiles of the Trappist-1 PPD (solid lines) superimposed onto those of the PPD taken from \cite{sch23} (dashed lines) after 0.01, 0.1, and 1 Myr of evolution.}
\label{fig:PTSig}
\end{figure}

Figure \ref{fig:PTSig} shows the surface density, midplane pressure, and temperature profiles derived from our Trappist-1 PPD model and from the PPD model used by \cite{sch23} after 0.01, 0.1, and 1 Myr of disk evolution. The comparison shows that the Trappist-1 PPD is cooler and less dense by a factor of $\sim$ 5--10 in the 0.5--10 AU region of the disk. The midplane pressure of the Trappist-1 PPD is also $\sim$30--70 times lower than in the PPD model of \cite{sch23}.

The evolution of the different water phases follows the advection-diffusion equation \citep{bir12,des17}:

\begin{equation}
\frac{\partial \Sigma_i}{\partial t} + \frac{1}{r} \frac{\partial}{\partial r} \left[ r \left( \Sigma_i v_i - D_i \Sigma_\mathrm{g} \frac{\partial}{\partial r} \left( \frac{\Sigma_i}{\Sigma_\mathrm{g}} \right) \right) \right] - \dot{Q}_i = 0,
\label{eq:advdiff}
\end{equation}

\noindent where  $\Sigma_i$ is the water surface density of phase $i$, either in vapor or ice form, $v_i$ and $D_i$ are the drift velocity and diffusivity, respectively, and $Q_i$ is the source or sink term of phase $i$, which takes the condensation and vaporization of water into account. The evolution of grains is governed by the two-population algorithm, which provides their mean drift speed and diffusivity \citep{bir12}. 

The water source and sink terms are governed by its equilibrium pressure \citep{wag11}:

\begin{equation}
\begin{split}
P_{\mathrm{eq}} = & P_{\mathrm{tp}} \exp{\left( \frac{T_{\mathrm{tp}}}{T} \right)} \left( -21.2144006 \left(\frac{T_{\mathrm{tp}}}{T}\right)^{1/300} \right. \\
&   +27.3203819 \left(\frac{T_{\mathrm{tp}}}{T}\right)^{2.10666667}  \\
& \left. -6.10598130 \left(\frac{T_{\mathrm{tp}}}{T}\right)^{1.70333333} \right),
\end{split}
\end{equation}

\noindent where $P_{\mathrm{tp}}$ and $T_{\mathrm{tp}}$ are the water triple point pressure (611.65 Pa) and temperature (273.16 K), respectively. The equilibrium pressure is compared with the partial pressure of water: 

\begin{equation}
P_\mathrm{v} = \frac{\Sigma_\mathrm{v} R T }{\sqrt{2\pi} \mu_{\ce{H2O}} H},
\end{equation}

\noindent where $\Sigma_\mathrm{v}$ is the water vapor surface density, $R$ is the perfect gas constant, $T$ is the disk temperature, $\mu_{\ce{H2O}}$ is the water molar mass, and $H$ is the disk scale height. If the vapor partial pressure is higher than the equilibrium one, then it condenses onto the pebbles and forms an icy shell. Condensation then results in a source term for water ice of
\begin{equation}
\Dot{Q}_{\mathrm{ice}} = \min{\left( \left(P_\mathrm{v} - P_{\mathrm{eq}} \right) \frac{2H \mu_{\ce{H2O}}}{RT \Delta t} ; \frac{\Sigma_{\mathrm{v},i}}{\Delta t} \right) }. 
\label{eq:cond}
\end{equation}

Conversely, if the vapor partial pressure is lower than the equilibrium pressure, ice vaporizes, resulting in a sink term for water ice of

\begin{equation}
\Dot{Q}_{\mathrm{ice}} =  - \min \left( \sqrt{\frac{8 \pi \mu_i}{RT}} \frac{3}{\pi \bar{a} \bar{\rho}} P_{\mathrm{eq}} \Sigma_{\mathrm{ice}} ; \frac{\Sigma_{\mathrm{ice}}}{\Delta t} \right),
\label{eq:ev}
\end{equation}

\noindent where $\bar{a}$ and $\bar{\rho}$ are the grains' mean size and mean density as given by the two-population algorithm, respectively.

We added a module that takes the dehydration of phyllosilicates into account. We considered that phyllosilicates dehydrate when they reach a temperature of 500 K in the disk \citep{mou23}.  Phyllosilicate-rich particles release water vapor at the PDL location in the PPD, which corresponds to the following source term for water vapor:

\begin{equation}
\Dot{Q} = \mathrm{WMF}_{\mathrm{phyllo}}  \frac{\Sigma_{\mathrm{p}}}{dt},
\end{equation}

\noindent where WMF$_{\ce{phyllo}}$ is the WMF in the phyllosilicates, which is varied in the 1--10\% range, and $\Sigma_\mathrm{p}$ is the surface density of phyllosilicates. We assumed that the phyllosilicate dehydration timescale is much shorter than the drift timescale \citep{cie05,mou23}. All the  water stored in the phyllosilicate dust or pebbles is released when they cross the PDL location. The PPD model was initialized with phyllosilicate grains of 0.1 $\mu$m in size. The initial phyllosilicate abundance, $x_\mathrm{p}^0 = \Sigma_\mathrm{p}^0/\Sigma_\mathrm{g}^0$, was set to $3.212 \times 10^{-3}$ \citep{lod09}.

\section{Results}

\label{sec:var_wmf}

\begin{figure*}[ht!]
\center 
\includegraphics[width=0.85\linewidth]{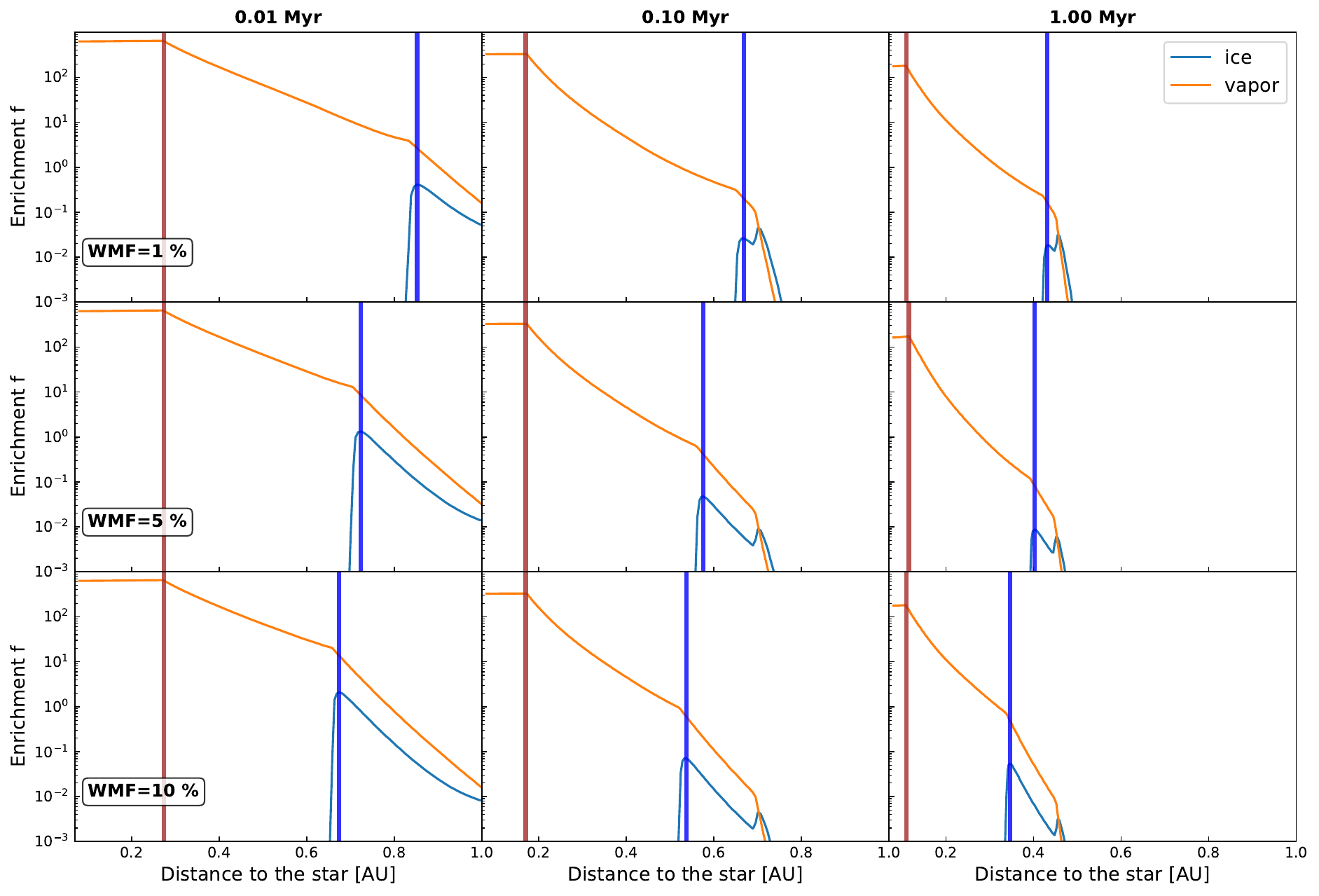}
\caption{From left to right: Enrichment profiles of water vapor (orange lines) and ice (blue lines) normalized with the initial abundance of water in phyllosilicates at $t$ = 0.01, 0.1, and 1 Myr. The initial WMFs in phyllosilicates here are 1\%, 5\%, and 10\%. The PDL and SL are represented as brown and blue bars, respectively.}
\label{fig:water_panel}
\end{figure*}

Figure \ref{fig:water_panel} displays water ice and vapor enrichment profiles in the PPD, with respect to the initial water abundance, after 0.01, 0.1, and 1 Myr of disk evolution. The initial water  abundance is defined as
\begin{equation}
   x_\mathrm{H_2O}^0 = \mathrm{WMF}_\mathrm{phyllo}~x_\mathrm{p}^0.
\end{equation}

\noindent Since the disk only contains phyllosilicate at $t$ = 0, the starting water abundance only depends on the water contained in phyllosilicate particles.  WMF$_{\ce{phyllo}}$ of 1\%, 5\%, and 10\% were used in the simulations.
The distance between the SL and the central star increases from 0.67 to 0.85 AU, 0.53 to 0.66 AU, and 0.35 to 0.43 AU when the WMF$_{\ce{phyllo}}$ value is ranged between 1\% and 10\% at 0.01, 0.1, and 1 Myr of PPD evolution, respectively.

The increase in this distance range is the direct consequence of the decrease in the amount of water vapor released from the pebbles into the PPD. Vapor accumulates at the PDL because its diffusion rate in the PPD is much lower than its production rate. A fraction of this vapor diffuses outward and crosses the SL.

Figure \ref{fig:water_panel} shows that the water ice enrichment reaches a maximum at $\sim$0.4, 1.3, and 2 times the initial water abundance before 0.01 Myr of PPD evolution, assuming a WMF$_{\ce{phylo}}$ of 1, 5, and 10\%, respectively. Between 0.01 and 0.1 Myr of PPD evolution, the water enrichment decreases by a factor of $\sim$20--30 down to 0.02, 0.04, and 0.07 times the initial water abundance. After 0.1 Myr, the ice enrichment remains almost constant up to 1 Myr, which corresponds to the end of our simulations.

\begin{figure}[h!]
\centering
\includegraphics[width=0.9\columnwidth]{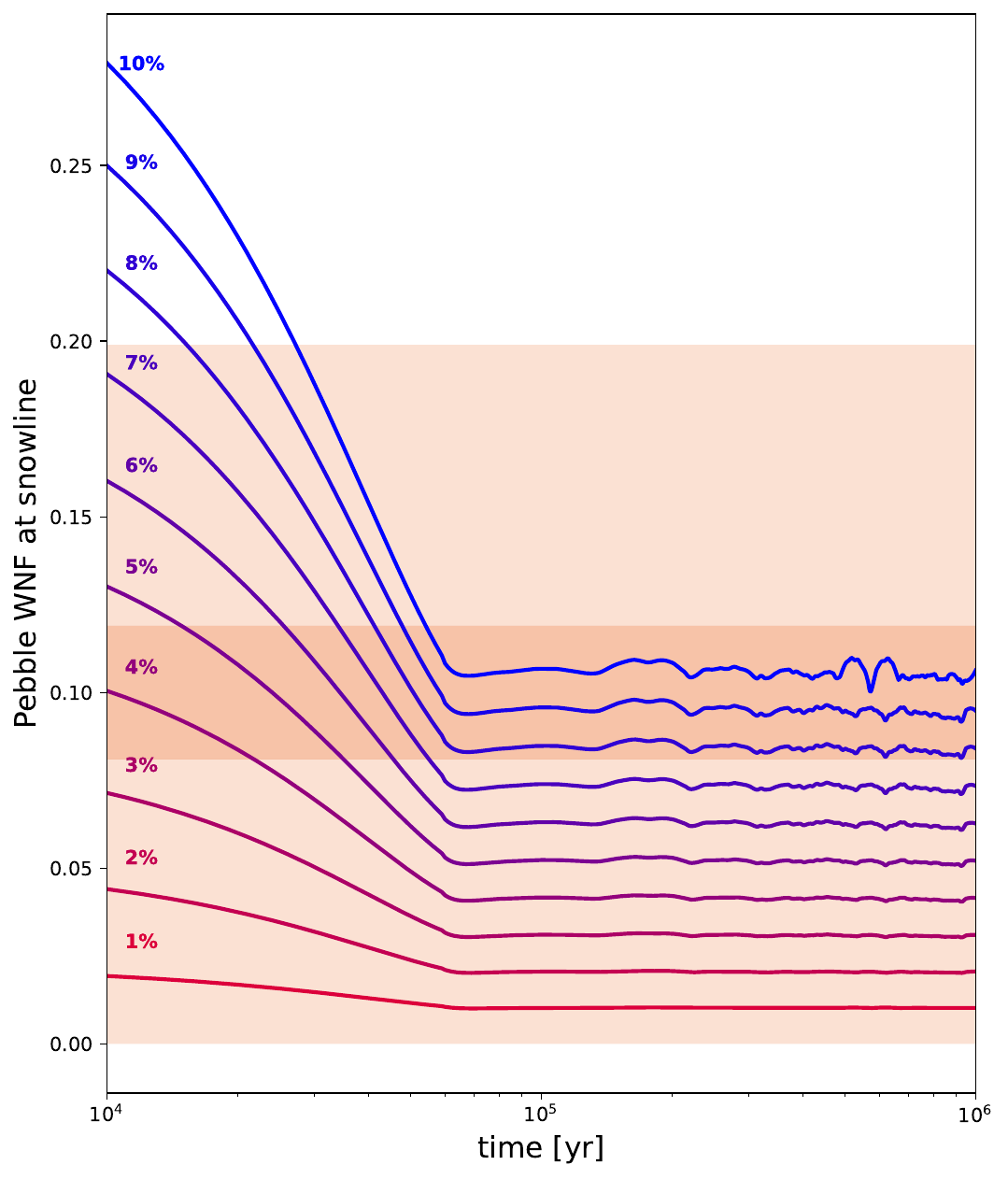}
\caption{Time evolution of the pebbles' WMFs at the SL location. The phyllosilicate WMF is varied between 1 and 10\%. The range of the WMF central values of the different Trappist-1 planets is bracketed by the dark orange box \citep{acu21}, and the 1$\sigma$ errors are shown in light orange.}
\label{fig:wmf_var}
\end{figure}

 Figure \ref{fig:wmf_var} represents the time evolution of the pebbles' WMF at the SL location when WMF$_\mathrm{phyllo}$ varies between 1 and 10\%. The dark orange area corresponds to the range of nominal WMFs (8.1--11.9\% ) computed by \cite{acu21} for Trappist-1 d, e, f, g, and h. The light orange area brackets the 1$\sigma$ error bars (0--19.9\%) associated with those measurements. The WMF of pebbles takes both the fraction of water in ice form and water trapped in phyllosilicates into account: 
 
\begin{equation}
\mathrm{WMF}= \frac{\mathrm{WMF}_\mathrm{phyllo}  \Sigma_\mathrm{p} + \Sigma_\mathrm{ice}}{  \Sigma_\mathrm{p} + \Sigma_\mathrm{ice}},
\end{equation}

\noindent with $\Sigma_\mathrm{ice}$ the surface density of ice in the PPD. Figure \ref{fig:wmf_var} shows that even in the absence of ice in the PPD at initialization, the pebbles' WMF is 2--3 times larger than WMF$_{\ce{phyllo}}$ after only 0.01 Myr of evolution. The pebbles' WMF ranges between 2 and 28\% when WMF$_{\ce{phyllo}}$ is varied between 1 and 10\%. As the PPD depletes, the amount of water vapor released from the phyllosilicates decreases, reducing the amount of ice condensed at the SL. Consequently, the pebbles' WMF converges toward WMF$_{\ce{phyllo}}$ after  0.07 Myr of PPD evolution.  This implies that only cases with WMF$_{\ce{phyllo}}$ in the 8--10\% range match the WMF of planets d--h after 0.07 Myr of PPD evolution. Interestingly, cases with pebbles that assume a WMF$_{\mathrm{phyllo}}$ in the 4--7\% range also match the WMF of planets d--h before 0.07 Myr of PPD evolution. However, after this epoch, water ice is too depleted in the PPD.

\section{Discussion}

The scenario we have explored for the Trappist-1 system divides the PPD into three distinct regions: (i) an inner region interior to the PDL in which solids consist of pebbles made of dehydrated silicates, (ii) an intermediary region located between the PDL and the SL populated with pebbles assembled from hydrated phyllosilicates, and (iii) a region located beyond the SL populated with pebbles consisting of hydrated phyllosilicate pebbles coated with water ice. 

In this scenario, the building blocks of planets b and c would have formed in regions interior to the PDL, while those of planets d--h would have formed at the SL. Planets b and c would have grown close to the star, possibly at their present-day location. On the other hand, planets d--h would have grown farther from the star than their current location \citep{orm17,col19,hua22,ogi22}. They could have migrated inward to their present-day location after formation. The inward migration of all planets could have led to the observed mean motion resonance chain formed by the seven planets \citep{orm17,bit19,col19,izi21}.

The dehydration temperature of phyllosilicates at pressure conditions relevant to those of PPDs is not well established. In our study we chose a dehydration temperature of 500 K to explore its effect on the pebbles' WMF \citep{mou23}. Varying the dehydration temperature between 400 K and 600 K has a significant impact on the timescales and extents of ice enrichment in the PPD. A higher dehydration temperature results in a PDL located closer to the star than in cases with lower temperatures. Therefore, colder dehydration temperatures reduce the distance between the PDL and SL, enhance the amount of vapor diffusing throughout the SL, and lead to higher abundances of solid water. A dehydration temperature of 400 K induces an ice enrichment up to $\sim$20 times the initial water abundance after 0.01 Myr of PPD evolution, assuming WMF$_{\ce{phyllo}}$ = 10\% in phyllosilicates. In contrast, only an ice enrichment of $\sim$0.5 is generated in the PPD, assuming a dehydration temperature of 600 K. These results show that our simulations are highly sensitive to the dehydration temperature. This variability impacts the evolution of the pebbles' WMF at the SL. 
Consequently, a lower dehydration temperature enables the formation of pebbles with WMFs consistent with those of planets d--h, assuming they start with WMF$_\mathrm{phyllo}$ lower than the nominal value (8\%) after 1 Myr of PPD evolution. Conversely, a higher dehydration temperature requires higher WMF$_\mathrm{phyllo}$ values to reproduce the WMFs of planets d--h after 1 Myr of PPD evolution.

To compare the sensitivity of our results against the PPD parameters, we reduced the PPD's initial mass from $10^{-1}$ to $10^{-2}$  $M_\star$. A less massive PPD has a lower density, resulting in reduced temperature and diffusion factor profiles. As a consequence, despite a reduced distance between the SL and the PDL of $\sim$0.3 AU, there is a decrease in the amount of vapor crossing the SL and condensing into ice. The water enrichment at the SL is reduced by a factor of $\sim$200 after 0.1 Myr of PPD evolution compared with the nominal enrichment, assuming a 10\%  WMF$_\mathrm{phyllo}$. After 1 Myr of PPD evolution, the disk dissipates almost completely and very few phyllosilicate grains remain, resulting in an even lower water vapor production. This indicates that the PPD has a somewhat significant mass (allowing it to generate pebbles) that is consistent with the WMFs of planets d--h. 

The proposed scenario relies on water loss by photolysis driven by FUV/XUV radiation from Trappist-1, whose intensity and time evolution remain unknown. If the FUV/XUV radiation levels were very low, it is likely that primordial water ice would remain abundant in the PPD. On the other hand, an extreme FUV/XUV radiation episode of sufficient duration would destroy the vapor released by phyllosilicate grains and pebbles, thereby preventing any ice formation at the SL. Consequently, a dry disk model represents an alternative scenario to the formation of the Trappist planets from volatile-rich material  \citep{orm17,bit19}. An observational test to investigate a dry disk scenario would be the discovery of a planet more distant than Trappist-1 h but with a density close to those of Trappist-1 b--c. This would imply that this planet formed from phyllosilicate-rich building blocks beyond the SL. In contrast, the same planet formed in a wet disk would have a density similar to those of planets d--h. Both scenarios could have been at play during the Trappist system formation. In this case, the inner planets would have formed from pebbles made of dehydrated phyllosilicates. The outer planets would have accreted from ice-rich pebbles sourced from (i) water vapor condensation after outward diffusion from the inner regions and (ii) pristine ice delivered from the outer regions. 

\begin{acknowledgements}

 The project leading to this publication has received funding from the Excellence Initiative of Aix-Marseille Universit\'e--A*Midex, a French ``Investissements d’Avenir program'' AMX-21-IET-018. This research holds as part of the project FACOM (ANR-22-CE49-0005-01\_ACT) and has benefited from a funding provided by l'Agence Nationale de la Recherche (ANR) under the Generic Call for Proposals 2022. We acknowledge Maxime Pineau for helpful discussions about phyllosilicate properties and experiments. 
\end{acknowledgements}

\end{document}